\documentclass[fleqn,usenatbib]{mnras}

\usepackage{newtxtext,newtxmath}
\usepackage[T1]{fontenc}
\usepackage{ae,aecompl}

\usepackage{graphicx}	% Including figure files
\usepackage{amsmath}	% Advanced maths commands
\usepackage{amssymb}	% Extra maths symbols
\usepackage{pifont}% http://ctan.org/pkg/pifont
\usepackage[normalem]{ulem}
\usepackage{footnote}
\newcommand{\cmark}{\ding{51}}%
\newcommand{\xmark}{\ding{55}}%

\title[A new approach to modelling GRB afterglows]{A new approach to modelling gamma-ray burst afterglows: Using Gaussian processes to account for the systematics}

\author[M. D. Aksulu et al.]{
M. D. Aksulu,$^{1}$\thanks{E-mail: m.d.aksulu@uva.nl}
R. A. M. J. Wijers,$^{1}$
H. J. van Eerten,$^{2}$
A. J. van der Horst$^{3,4}$
\\
$^{1}$  Anton Pannekoek Institute for Astronomy, University of Amsterdam, Science Park 904, NL-1098 XH Amsterdam, the Netherlands\\
$^{2}$ Department of Physics, University of Bath, Claverton Down, Bath BA2 7AY, UK\\
$^{3}$ Department of Physics, The George Washington University, 725 21st Street NW, Washington, DC 20052, USA\\
$^{4}$ Astronomy, Physics, and Statistics Institute of Sciences (APSIS), 725 21st Street NW, Washington, DC 20052, USA\\
}

\date{Accepted XXX. Received YYY; in original form ZZZ}

\pubyear{2015}

\begin{document}
\label{firstpage}
\pagerange{\pageref{firstpage}--\pageref{lastpage}}
\maketitle

%@arxiver{970508-lc_gp_mean_radio.pdf, E_K_iso_theta_0-alt-3-sigma.pdf}
\begin{abstract}
The afterglow emission from gamma-ray bursts (GRBs) is a valuable source of information to understand the physics of these energetic explosions. The fireball model has become the standard to describe the evolution of the afterglow emission over time and frequency. Thanks to recent developments in the theory of afterglows and numerical simulations of relativistic outflows, we are able to model the afterglow emission with realistic dynamics and radiative processes. Although the models agree with observations remarkably well, the afterglow emission still contains additional physics, instrumental systematics, and propagation effects which make the modelling of these events challenging. In this work, we present a new approach to modelling GRB afterglows, using Gaussian processes (GPs) to take into account systematics in the afterglow data. We show that, using this new approach, it is possible to obtain more reliable estimates of the explosion and microphysical parameters of GRBs. We present fit results for 5 long GRBs and find a preliminary correlation between the isotropic energetics and opening angles of GRBs, which confirms the idea of a common energy reservoir for the kinetic energy of long GRBs.
\end{abstract}

\begin{keywords}
gamma-ray burst: general -- gamma-ray burst: individual: GRB 970508, 980703, 990510, 991208, 991216 -- methods: data analysis -- methods: statistical
\end{keywords}

%%%%%%%%%%%%%%%%%%%%%%%%%%%%%%%%%%%%%%%%%%%%%%%%%%

%%%%%%%%%%%%%%%%% BODY OF PAPER %%%%%%%%%%%%%%%%%%

\section{Introduction}
Gamma-ray bursts (GRBs) are the most energetic explosions in the Universe. They are either the result of the collapse of massive stars (long GRBs) \citep{Woosley1993}, or of compact object mergers where at least one of the objects is a neutron star (short GRBs) \citep{Eichler1989}; for a review see, e.g. \cite{Piran2004}. During these catastrophic events, an ultra-relativistic, collimated outflow is generated by a compact central engine \citep{Rees1992}. Initially, GRBs are detected as prompt $\gamma$-ray flashes. The exact emission mechanism which produces these $\gamma$ rays is still debated; for a review see, e.g. \cite{Kumar2015}. As the outflow starts to interact with the circumburst medium, it starts to decelerate and forms a relativistic, collisonless shock where charged particles are accelerated in tangled magnetic fields and emit synchrotron emission across the whole electromagnetic spectrum \citep{Rees1992}. This emission is called the afterglow of the GRB. It is possible to understand more about the physics of GRBs by modelling the afterglow. The afterglow emission reveals how the dynamics of such relativistic shocks evolve over time as well as the microphysical properties in such extreme acceleration regions \citep{Wijers1997, Sari1998, Wijers1999, Panaitescu2002, Yost2003}.

With the launch of the \textit{Neil Gehrels Swift Observatory} \citep{gehrels04}, the detection rate of GRB afterglows has significantly increased. Together with multi-wavelength ground and space based follow-up, and the start of the multi-messenger era, we now have a wealth of data on GRB afterglows \citep{abbott17a, MAGIC2019}. Moreover, recent advancements in afterglow theory and numerical hydrodynamics allow us to model the dynamics and emission mechanism of GRB afterglows much more reliably \citep{vanEerten2018}. Although the models agree with the general trends of the afterglow data, it is still challenging to get reliable estimates of GRB parameters because of additional physics which is not included in the models (e.g. self-synchrotron Compton scattering effects, reverse shock emission), instrumental systematics, and propagation effects (e.g. scintillation in radio, absorption by the host galaxy gas and dust in the optical and X-ray regimes). All these effects introduce systematic deviations to the afterglow observations, and result in a more complex flux evolution over time and frequency than predicted by the models. \cite{Gompertz2018} have shown that there must be intrinsic errors involved when modelling GRB afterglows, by using closure relations to show that the data exhibits inconsistencies which cannot be explained by the measurement errors. In this work, we show that systematic deviations put unrealistically tight constraints on the model parameters when performing parameter estimation where the likelihood function is only proportional to the $\chi^2$ value. 

In this paper we introduce a new approach to fitting GRB afterglow data, by modelling the systematics using Gaussian processes (GPs). This way, the model parameters are not bound by artifacts of systematic deviations, and Bayesian parameter estimation gives more reliable parameter uncertainties. In Section~\ref{sec:method} we explain the method in detail, in Section~\ref{sec:test} we present test results with synthetic data sets, and compare to results obtained by conventional modelling. Moreover, in Section~\ref{sec:res} we apply this method to 5 long GRB afterglow data sets and present the results. In Section~\ref{sec:discuss} we further elaborate on the modelling of GRB afterglows, and we conclude in Section~\ref{sec:conclude}.

\section{Method}
\label{sec:method}
GPs are a generalization of the Gaussian probability distribution, in the sense that, GPs enable us to define a probability distribution over functions instead of variables or vectors \citep{rasmussen}. GPs are non-parametric, stochastic processes, and are therefore a useful tool in regression problems where the underlying model of the data is unknown, as is the case for the systematic differences between GRB afterglow models and observations. 

In this section we describe a Gaussian process framework for modelling the systematics in the GRB afterglow data sets. We follow the same methodology described in \cite{Gibson2012}, where they used the same approach to model transit light curves of exoplanets, which are affected by significant systematics.

\subsection{GRB afterglow data}
In order to solve for the many GRB afterglow model parameters, a well-sampled multi-wavelength data set is required. The data set consists of $N$ flux measurements, $\boldsymbol{y} = ({f_\nu}_1, \dots, {f_\nu}_N)^T$, measured at times and frequencies $\boldsymbol{\mathrm{X}} = (\boldsymbol{x}_1, \dots, \boldsymbol{x}_N)^T = ((t_1, \nu_1)^T, \dots, (t_N, \nu_N)^T)^T$, where $\boldsymbol{\mathrm{X}}$ is an $N\times2$ matrix. The reported uncertainties of the flux measurements are expressed as $\boldsymbol{\sigma} = (\sigma_1, \dots, \sigma_N)^T$.

\subsection{Gaussian process framework}

In this work, we use a GP model to take into account any possible systematics in the GRB afterglow data in a non-parametric fashion, where the systematics are described as,

\begin{equation}
    f\left(t, \nu\right) \sim \mathcal{GP}\left(\mu(t, \nu, \boldsymbol{\phi}), \boldsymbol{\Sigma}(t, \nu, \boldsymbol{\theta})\right),
\end{equation}
\noindent where $\mu$ is the mean function of the GP (i.e. the afterglow model), $\boldsymbol{\Sigma}$ is the covariance matrix of the GP model, $\boldsymbol{\phi}$ and $\boldsymbol{\theta}$ represent the GRB parameters and the, so called, \textit{hyperparameters} of the GP model respectively.

The log likelihood of the GP model is described as,
\begin{equation}
    \log\mathcal{L(\boldsymbol{r}|\boldsymbol{\mathrm{X}}, \boldsymbol{\theta}, \boldsymbol{\phi})} = -\frac{1}{2}\boldsymbol{r}^T\boldsymbol{\Sigma}^{-1}\boldsymbol{r} - \frac{1}{2}\log|\boldsymbol{\Sigma}| - \frac{N}{2}\log(2\pi)
    \label{eq:gp_likelihood}
\end{equation}
\noindent where $\boldsymbol{r}$ is the residual of the afterglow model with respect to the observed flux density values. We define the residual as,
\begin{equation}
    \boldsymbol{r} = \log\boldsymbol{y} - \log\boldsymbol{\mu},
\end{equation}
\noindent due to the fact that the measured flux densities vary over orders of magnitudes with time and frequency. In such cases it is common to model the logarithm of the measured values \citep{Snelson2004}. Therefore, we exclude any negative flux measurements from the data sets when modelling.

The covariance matrix, $\boldsymbol{\Sigma}$, defines how correlated the data points are over observer time and frequency. In order to construct the covariance matrix, a squared-exponential kernel is chosen, over the 2D input space (time and frequency),

\begin{equation}
    \boldsymbol{\Sigma}_{ij} = k(\boldsymbol{\mathrm{X}}_i, \boldsymbol{\mathrm{X}}_j) = A \exp\left[-\frac{1}{2}\sum^2_{k = 1}\frac{(\boldsymbol{\mathrm{X}}_{ik}- \boldsymbol{\mathrm{X}}_{jk})^2}{l_k^2}\right] + \delta_{ij}\sigma_w^2,
\end{equation}

\noindent where $A$ represents the amplitude of the correlations,  $l_1$ and $l_2$ determine the length scales of the correlations over time and frequency respectively, and $\sigma_w$ represents the amount of white noise in the data set. These parameters are called the hyperparameters of the GP and need to be marginalized together with the model parameters. The white noise parameter is formulated as $\sigma_w = \sigma_{\log f_\nu} \sigma_h$, where $\sigma_{\log f_\nu}$ is the uncertainty in the logarithm of the flux measurements and $\sigma_h$ is the hyperparameter which scales the reported uncertainties. Thus, the hyperparameters can be expressed as,

\begin{equation}
    \boldsymbol{\theta} = (A, l_1, l_2, \sigma_h)^T.
    \label{eq:hyperparams}
\end{equation}
 
In this work, we use the \texttt{george} Python package \citep{george} as the GP framework. \texttt{george} enables us to calculate efficiently the covariance matrix and the GP likelihood even for relatively large data sets and it is designed to be used with any external optimization/sampling algorithm.

\subsection{Model}
\label{sec:model}
We assume a relativistic, collimated, outflow interacting with the circumburst medium (CBM), forming a pair of shocks propagating into the ejecta (short-lived reverse shock) and into the CBM (long-lived forward shock) where charged particles are accelerated and emit synchrotron radiation \citep{Sari1998, Wijers1999, Granot2002}. In this work we only consider the emission originating from the forward shock. The forward shock model has been able to successfully describe the spectral and temporal evolution of GRB afterglows.

In this work, we incorporate \texttt{scalefit} \citep[][Ryan et al. in prep.]{Ryan2015}, as the mean function of the GP model. \texttt{scalefit} is an afterglow model, which makes use of pre-calculated tables of spectral features (i.e self-absorption break $\nu_a$, injection break $\nu_m$, cooling break $\nu_c$, and peak flux density of the spectrum $f_{\nu, \mathrm{peak}}$) over decades in time and for different observing angles. \texttt{scalefit} takes advantage of scale invariance to calculate the observed flux density for given explosion and microphysical parameters, observer times and frequencies. The model parameters are described as,

\begin{equation}
    \boldsymbol{\phi} = (\theta_0, E_{K, \mathrm{iso}}, n_0, \theta_{\mathrm{obs}}, p, \epsilon_B, \Bar{\epsilon}_e, \xi_N)^T,
    \label{eq:model_params}
\end{equation}

\noindent where $\theta_0$ is the opening angle of the jet, $E_{K, \mathrm{iso}}$ is the isotropic-equivalent kinetic energy of the explosion, $n_0$ is the circumburst number density, $\theta_{\mathrm{obs}}$ is the observing angle, $p$ is the power-law index of the accelerated electron population, $\epsilon_B$ is the fraction of thermal energy in the magnetic fields, $\Bar{\epsilon}_e \equiv \frac{p - 2}{p - 1}\epsilon_e$ where $\epsilon_e$ is the fraction of thermal energy in the accelerated electrons, and $\xi_N$ is the fraction of electrons being accelerated. 

\texttt{boxfit} \citep{boxfit} is used to produce the tables containing the spectral features. \texttt{boxfit} is a GRB afterglow modelling tool, which makes use of pre-calculated hydrodynamics data and solves radiative transfer equations during runtime. Since it relies on hydrodynamics data, it is able to model the dynamics of the blast wave reliably. \texttt{boxfit} has been used to successfully model the broadband emission from various afterglows (see e.g. \citealt{Guidorzi2014, Higgins2019, Kangas2020}), but its computationally expensive repeated radiative transfer calculations are a drawback when implementing \texttt{boxfit} in a sampling algorithm. \texttt{scalefit} instead draws from a table that reproduces the spectral breaks and peak fluxes from boxfit exactly, but approximates the spectral curvature across breaks when reconstructing spectra. Through its approximation of spectral curvature, \texttt{scalefit} avoids the need for repeated radiative transfer calculations and allows for fast computation (for applications, see e.g. \citealt{Ryan2015, Zhang2015}). This offers a good compromise between speed and accuracy. In this work, aimed at GPs, we make use of the fact that due to the slight differences in their approach the different modelling tools produce afterglow light curves with different relative systematics, leaving a detailed comparison of the relative merit between the methods for afterglow modelling in general for future work.

\subsection{Regression}
\label{sec:reg}
In order to marginalize over the model parameters and the hyperparameters of the GP, we make use of \texttt{pymultinest} \citep{pymultinest}, which is the Python implementation of the MultiNest nested sampling algorithm \citep{multinest}. Sampling from complex objective functions can be challenging as algorithms can get stuck in local maxima. The main advantage of using \texttt{pymultinest} is that it is able to converge on the global maximum with high efficiency (i.e. relatively small number of function evaluations). For all the presented results, \texttt{pymultinest} is used in the importance sampling mode \citep{Feroz2019} with mode separation disabled. We use 1000 initial live points and use an evidence tolerance of 0.5 as our convergence criterion. These values are adapted from \cite{multinest}.

The fraction of accelerated electrons, $\xi_N$, is degenerate with respect to  $(E_{K,\mathrm{iso}}, n_0, \epsilon_B, \Bar{\epsilon}_e)$, where $(E_{K,\mathrm{iso}}, n_0)$ are proportional to $1 / \xi_N$, and $(\epsilon_B,  \bar{\epsilon}_e)$ are proportional to $\xi_N$ \citep{Eichler2005}. Because of this degeneracy, we fix $\xi_N$ to be 0.10. Canonically, $\xi_N$ is set to unity when modelling GRB afterglows, however, for our sample we find that a smaller value for $\xi_N$ gives more physical results for $\epsilon_B$ and $\Bar{\epsilon}_e$, since accepting the canonical value results in non-physical parameter values (e.g. $\epsilon_B + \epsilon_e > 1$). Moreover, particle-in-cell simulations have shown that $\xi_N$ can be as low as 0.01, depending on the shock conditions \citep{Sironi2011}.

\begin{figure}
 \includegraphics[width=\columnwidth]{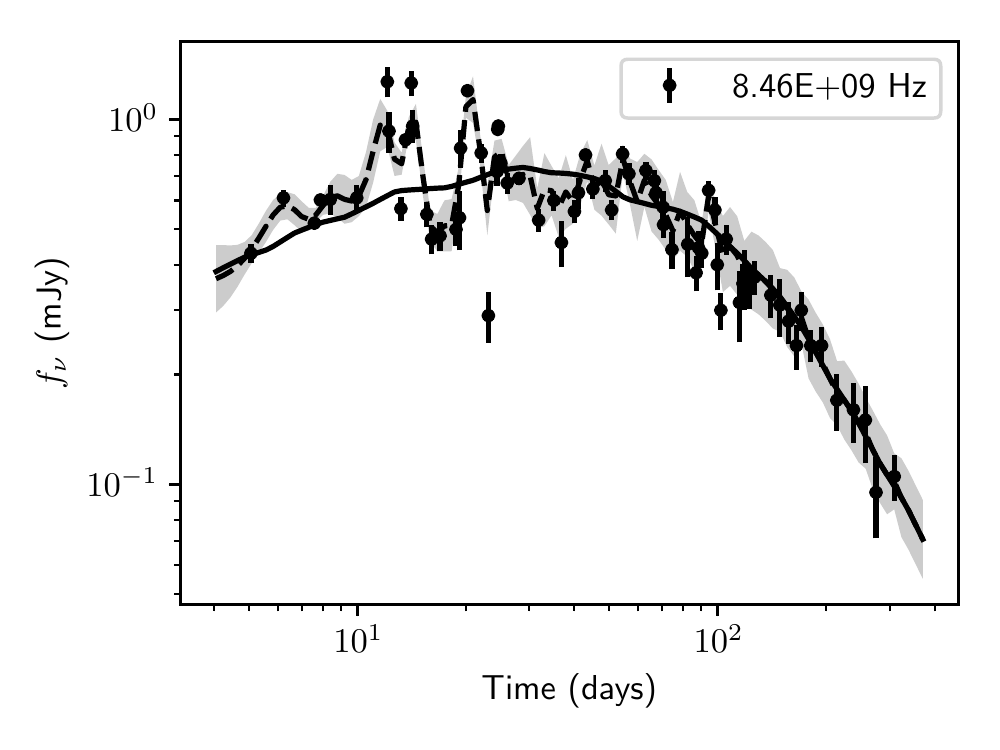}
 \caption{Example regression result from modelling GRB 970508 using the GP model. The radio light curve at 8.46 GHz is shown, where the solid line is the \texttt{scalefit} light curve, the dashed line is the mean predicted by the GP model and the shaded area represents the 1-$\sigma$ uncertainty of the GP model. It can be seen that at early times the data is heavily affected by scintillation, and the systematics are modelled by the GP. The variability in the model (solid line) is due to numerical noise.}
 \label{fig:lc_gp_mean_970508_radio}
\end{figure}

Regression is performed by marginalizing over both the hyperparameters and model parameters (see Equations \ref{eq:hyperparams} and \ref{eq:model_params}). In all of the fits presented in this work, we assume that the systematics are uncorrelated over the frequency domain by fixing the hyperparameter $l_2$ to a very small number. We recognize that this assumption may not hold for regions of the spectrum where the frequency domain is sampled closely (e.g. radio or optical observations at similar frequencies). Also, dust extinction may lead to correlated noise in the frequency space over several decades. However, after correcting for dust extinction, when the data set spans over multiple decades in frequency, the emission in radio, optical and X-rays will not be correlated. In Figure~\ref{fig:lc_gp_mean_970508_radio}, we present an example regression result for the radio light curve of GRB 970508, which contains significant variability in radio bands at early times due to interstellar scintillation. It can be seen that over time, as the shock front expands and the source size increases, the variability decreases.

\section{Application to synthetic data}
\label{sec:test}
In order to test the effectiveness of the proposed method, we generate synthetic data sets and try to recover the true parameters by modelling the synthetic data using both the conventional method of sampling the $\chi^2$ likelihood and the proposed method of sampling the GP log likelihood function (Equation \ref{eq:gp_likelihood}). 

Two sets of synthetic data are generated using \texttt{scalefit} (Model 1 from now on) and \texttt{boxfit} (Model 2 from now on) as the underlying model. The synthetic data sets are generated in radio, optical and X-ray bands and across 10 time epochs, which are log-uniformly separated. The uncertainty fractions are chosen to be 10\%, 2\%, and 10\%, for radio, optical and X-ray bands respectively. The data points are generated by drawing from a Gaussian distribution with the model value (either Model 1 or 2) as the mean and the corresponding uncertainty as the standard deviation.

% As explained in Section \ref{sec:model}, \texttt{scalefit} is an approximation of \texttt{boxfit}. Therefore at certain time epochs, frequencies and parameter values, Model 1 and 2 may predict different flux densities.
In this work, we model any type of data set using Model 1 (see Section~\ref{sec:model}). Therefore, the synthetic data set generated with Model 1 contains only white noise, whereas the synthetic data set generated with Model 2 also contains systematics with respect to Model 1. This allows us to test the performance of the GP model, both in the absence and presence of systematic differences. For all the synthetic data modelling, we use the same prior for the parameters, which is presented in Table~\ref{tab:prior_synth}. We select fiducial GRB parameter values for our synthetic data sets; 
$(\theta_0, E_{K, \mathrm{iso}}, n_0, \theta_{\mathrm{obs}} / \theta_0, p, \epsilon_B, \epsilon_e) = (0.10, 10^{53}, 1.00, 0.30, 2.4, 10^{-2}, 10^{-1})$. 

\begin{table}
    \setlength{\tabcolsep}{5pt}
    \def\arraystretch{1.25}
    \caption{Assumed priors for modelling synthetic data sets.}
    \begin{tabular}{l|r}
    \hline
    \multicolumn{1}{l}{Parameter range} & \multicolumn{1}{r}{Prior distribution} \\
    \hline
    $0.01<\theta_0<1.6$ & log-uniform\\
    
    $10^{50}<E_{K, \mathrm{iso}}<10^{55}$ & log-uniform\\
    
    $10^{-4}<n_{0}<1000$ &log-uniform\\
    
    $0<\theta_{\mathrm{obs}} / \theta_0<2$ & uniform\\
    
    $1.5<p<3.0$& uniform\\
    
    $10^{-7}<\epsilon_B<0.50$& log-uniform\\
    
    $10^{-4}<\bar{\epsilon_e}<10$& log-uniform\\
    \hline
    \end{tabular}
    \label{tab:prior_synth}
\end{table}

\begin{table}
    \caption{Fit results for the synthetic data set generated using Model 1. The data set contains only white noise as described in Section~\ref{sec:test}.  Results from both $\chi^2$ and GP ($\mathcal{GP}$) likelihood sampling are presented. All the uncertainties on the parameters represent the 95\% credible interval. Parameter estimations which include and exclude the true parameter value within the 95\% credible interval are marked as \cmark and \xmark, respectively. }
    \label{tab:synth_scalefit_res}
    \setlength{\tabcolsep}{5pt}
    \def\arraystretch{1.25}
    \begin{tabular} {l r r r}
        \hline
        Parameter &  \multicolumn{1}{c}{$\chi^2$} & \multicolumn{1}{c}{$\mathcal{GP}$} & True value\\
        \hline
        {\boldmath$\theta_0$} & {$0.0967^{+0.016}_{-0.0061}$ \cmark} & {$0.098^{+0.017}_{-0.029}$ \cmark} & 0.10\\
        
        {\boldmath$\log_{10}(E_{K, \mathrm{iso}, 53})$} & {$0.01^{+0.17}_{-0.21}$ \cmark} & {$-0.04^{+0.23}_{-0.24}$ \cmark} & 0.00\\
        
        {\boldmath$\log_{10}(n_0) $} & {$-0.09^{+0.18}_{-0.23}$ \cmark} & {$-0.12^{+0.26}_{-0.32}$ \cmark} & 0.00\\
        
        {\boldmath$\theta_{\mathrm{obs}}/\theta_0$} & {$0.29^{+0.17}_{-0.13}$ \cmark} & {$0.27^{+0.26}_{-0.22}$ \cmark} & 0.30\\
        
        {\boldmath$p              $} & {$2.405^{+0.022}_{-0.018}$ \cmark} & {$2.406^{+0.028}_{-0.042}$ \cmark} & 2.40\\
        
        {\boldmath$\log_{10}(\epsilon_B)$} & {$-1.94^{+0.27}_{-0.24}$ \cmark} & {$-1.91^{+0.37}_{-0.34}$ \cmark} & -2.00\\
        
        {\boldmath$\log_{10}(\bar{\epsilon}_e)$} & {$-1.48^{+0.15}_{-0.20}$ \cmark} & {$-1.51^{+0.16}_{-0.15}$ \cmark} & -1.54\\
        \hline
    \end{tabular}
\end{table}

\begin{table}
    \caption{Fit results for the synthetic data set generated using Model 2. The data set contains both white noise and systematics as described in Section~\ref{sec:test}. Results from both $\chi^2$ and GP ($\mathcal{GP}$) likelihood sampling are presented. All the uncertainties on the parameters represent the 95\% credible interval. Parameter estimations which include and exclude the true parameter value within the 95\% credible interval are marked as \cmark and \xmark, respectively. }
    \label{tab:synth_boxfit_res}
    \setlength{\tabcolsep}{5pt}
    \def\arraystretch{1.25}
    \begin{tabular} {l r r r}
        \hline
        Parameter &  \multicolumn{1}{c}{$\chi^2$} & \multicolumn{1}{c}{$\mathcal{GP}$} & True value\\
        \hline
        {\boldmath$\theta_0       $} & $0.03570^{+0.00081}_{-0.00072}$ \xmark & {$0.071^{+0.062}_{-0.039}$ \cmark} & 0.10\\
        
        {\boldmath$\log_{10}(E_{K, \mathrm{iso}, 53})$} & {$0.519^{+0.034}_{-0.034}   $ \xmark} & {$0.57^{+0.87}_{-1.0}$ \cmark} & 0.00\\
        
        {\boldmath$\log_{10}(n_0) $} & {$-2.669^{+0.044}_{-0.048} $ \xmark} & {$-0.2^{+1.2}_{-1.3}$ \cmark} & 0.00\\
        
        {\boldmath$\theta_{\mathrm{obs}}/\theta_0$} & {$0.641^{+0.033}_{-0.038}   $ \xmark} & {$0.70^{+0.46}_{-0.58}$ \cmark} & 0.30\\
        
        {\boldmath$p              $} & {$2.469^{+0.014}_{-0.014}$ \xmark} & {$2.33^{+0.11}_{-0.12}$ \cmark} & 2.40\\
        
        {\boldmath$\log_{10}(\epsilon_B)$} & {$-0.538^{+0.081}_{-0.081}$ \xmark} & {$-2.0^{+1.5}_{-1.6}$ \cmark} & -2.00\\
        
        {\boldmath$\log_{10}(\bar{\epsilon}_e)$} & {$-1.779^{+0.018}_{-0.017}$ \xmark} & {$-1.96^{+0.46}_{-0.52}$ \cmark} & -1.54\\
        \hline
    \end{tabular}
\end{table}

Tables \ref{tab:synth_scalefit_res} and \ref{tab:synth_boxfit_res} show fit results for both data sets and modelling approaches. As it can be seen in Table~\ref{tab:synth_scalefit_res}, the GP model and $\chi^2$ sampling perform similarly in the absence of systematics. Overall, the GP model results in larger parameter uncertainties. When there are systematics involved, the shortcomings of the $\chi^2$ sampling approach stand out. In Table~\ref{tab:synth_boxfit_res}, we show that the $\chi^2$ sampling technique is unable to recover any of the true parameters, despite inferring small uncertainties on the parameters. On the other hand the GP model is able to recover every parameter within the 95\% credible interval.

\begin{figure}
 \includegraphics[width=\columnwidth]{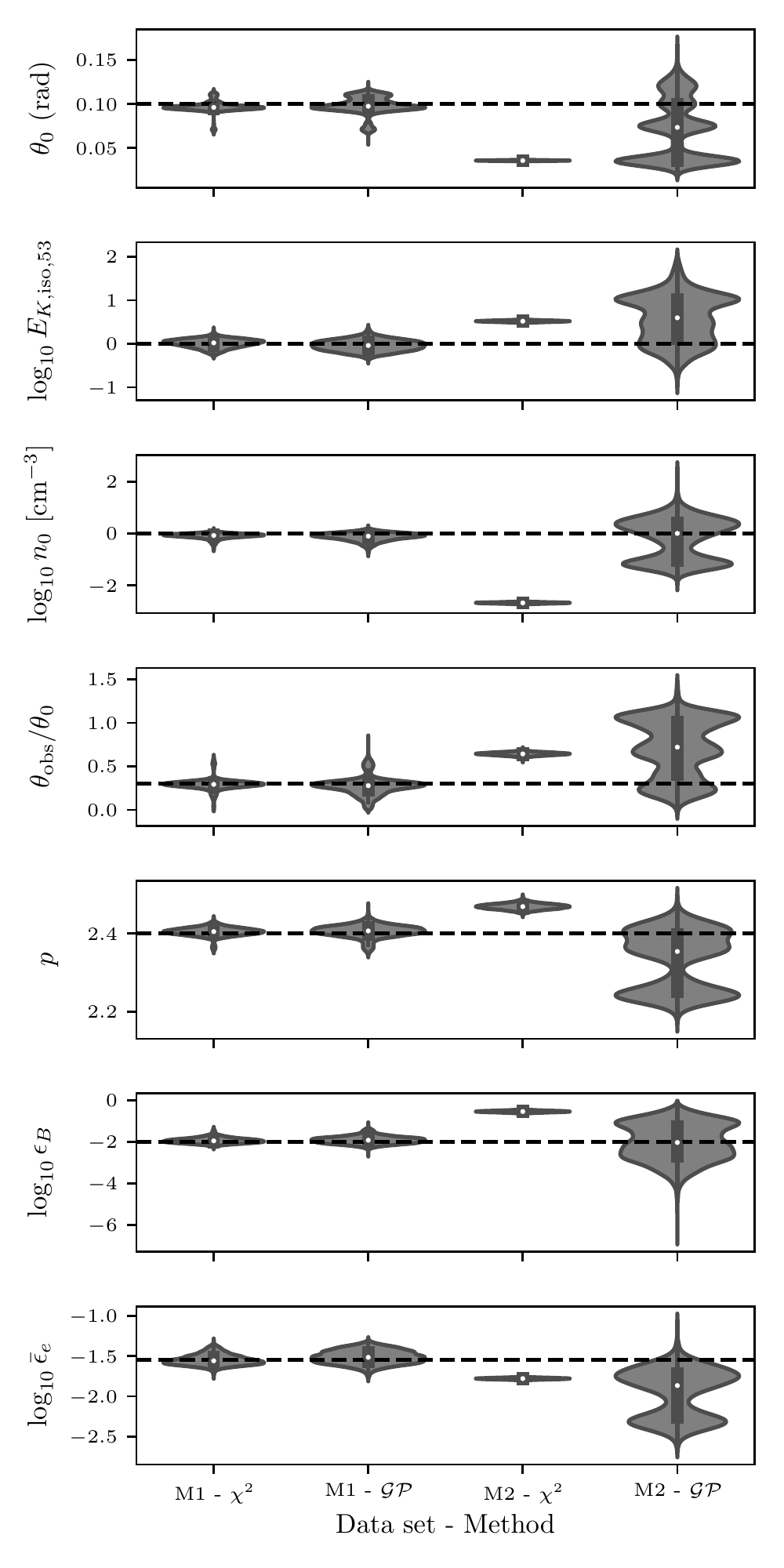}
 \caption{Violin plot of the fit results for different synthetic data sets and methods. Data sets which are generated by Model 2 (M2) contain systematics and white noise. Data sets generated with Model 1 (M1) contain no systematics but the same amount of white noise as Model 2. Both data sets are fitted using $\chi^2$ sampling ($\chi^2$) and sampling the GP likelihood in Equation \ref{eq:gp_likelihood} ($\mathcal{GP}$). The horizontal dashed lines represent the true parameter values.}
 \label{fig:test_violin}
\end{figure}

In Figure~\ref{fig:test_violin} we show the fit results for all synthetic data sets and modelling approaches in the form of violin plots. Violin plots are a way to visualize the marginalized distributions of parameters in a compact way, where the shaded area represents the normalized histogram of the posterior samples, and the solid bar shows the interquartile range of the distribution. As it can be seen, the GP model results in larger uncertainties, more complex marginal distributions, and more accurate parameter estimations.

In order to investigate further whether sampling the GP likelihood results in more reliable parameter inferences, regardless of the chosen parameters for the synthetic data sets, we generate 100 sets of synthetic data both using Model 1 and Model 2 with randomly chosen GRB parameters. We fit all of the data sets using both the proposed method of GP likelihood sampling and $\chi^2$ sampling. We perform coverage measurements on these fit results to determine how accurate the inferred uncertainties are. Coverage measurements are performed by fitting 100 synthetic data sets and counting how many times the true parameter was recovered for a given confidence region. 

\begin{figure*}
 \includegraphics[width=\textwidth]{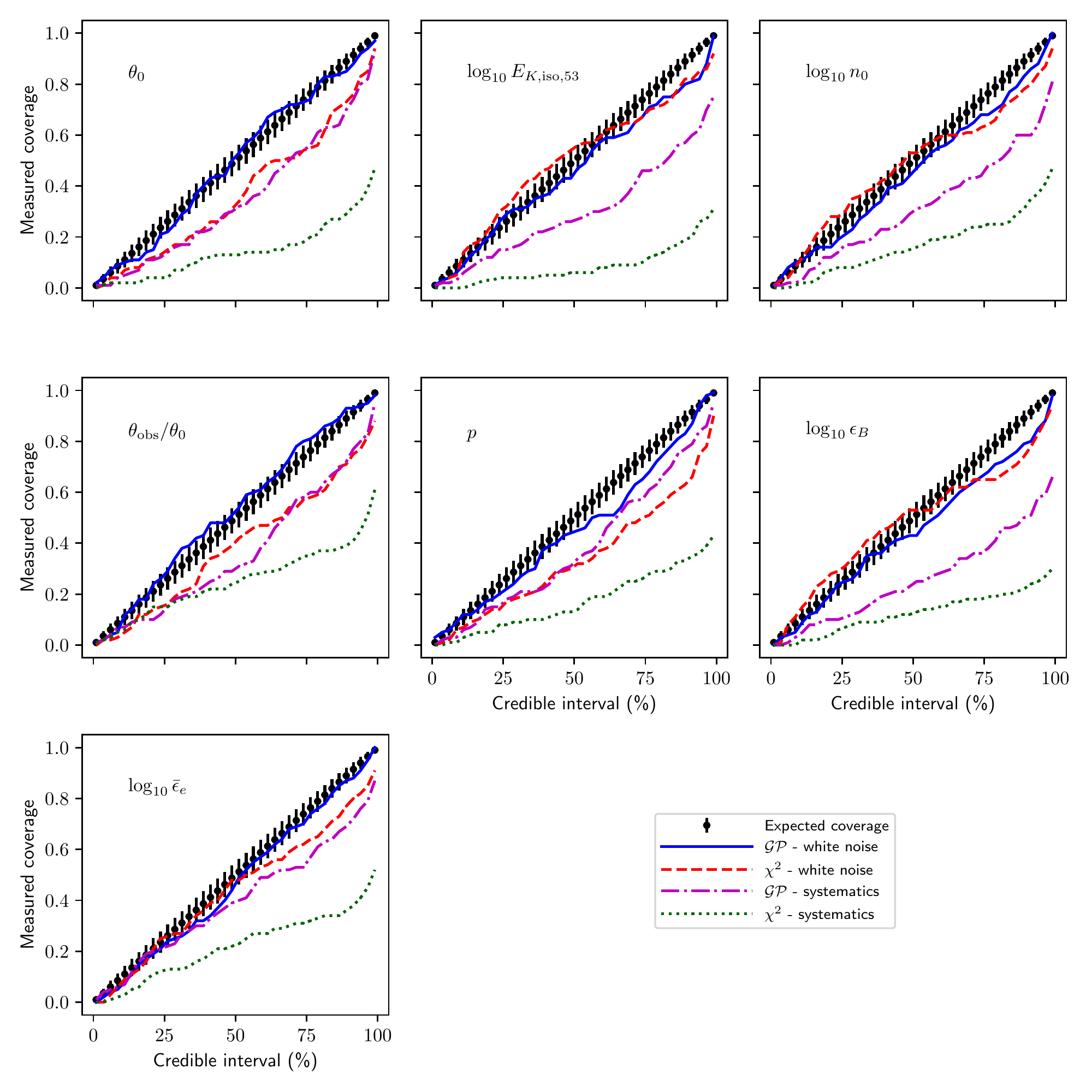}
 \caption{Coverage measurement results for GRB parameters. The blue and red lines show the coverage measurement results in the case where the data only contains white noise (synthetic data set generated by Model 1) for GP model regression ($\mathcal{GP}$) and $\chi^2$ sampling, respectively. The magenta and yellow lines show the coverage measurement results in the case where the data contains systematic deviations (synthetic data set generated by Model 2) for GP model regression and $\chi^2$ sampling, respectively. The black points show the ideally expected coverage, where the error bars are calculated using the binomial uncertainty $\sigma = \sqrt{p (1 - p) / N}$, where $p$ is the probability of containing the true parameter (credible interval) and $N$ is the number of samples (100 in our case).}
 \label{fig:cm}
\end{figure*}

In Figure~\ref{fig:cm}, we show the coverage measurement results for each GRB parameter both in the presence and absence of systematic deviations. These plots show the fraction of successfully recovered parameters (vertical axis) for a given credible interval \footnote{Credible interval is the Bayesian analogue of confidence interval.} (horizontal axis) \citep{Sellentin2019}. The black points show the ideally expected coverage, where the error bars are calculated using the binomial uncertainty, given by,

\begin{equation}
    \sigma = \sqrt{p (1 - p) / N}
\end{equation}
\noindent where $p$ is the probability of containing the true parameter (credible interval) and $N$ is the number of samples (100 in our case). The coverage measurements show that the GP model performs better both in the presence and absence of systematic deviations, as the measured coverage for the GP model is closer to the ideal case. $\chi^2$ sampling underestimates the errors on the parameters, especially for parameters which affect the temporal slope of the light curves ($\theta_0$, $\theta_{\mathrm{obs}}$ and $p$). In the presence of systematic deviations, even the GP model underestimates the errors on the parameters, however, less so than $\chi^2$ sampling.

\section{Application to archival GRB afterglow data}
\label{sec:res}
\begin{figure*}
 \includegraphics[width=\textwidth]{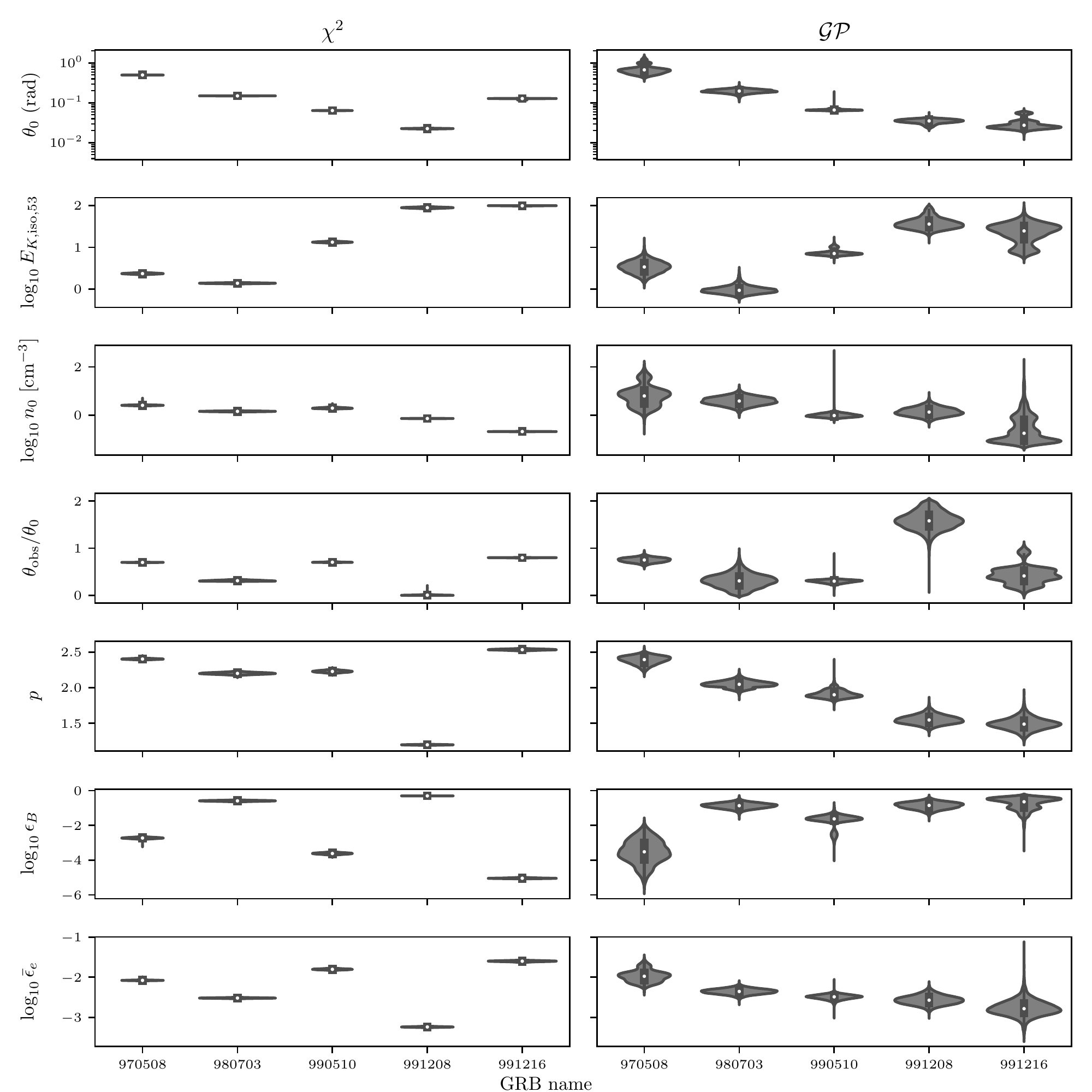}
 \caption{Violin plot showing the inferred marginalized distributions of the GRB parameters for the long GRB sample. The panel on the left shows the results for $\chi^2$ sampling whereas the panel on the right shows the results from GP likelihood sampling.}
 \label{fig:res_violin}
\end{figure*}

In this section we present fit results for 5 long GRB afterglows for which significant modeling has already been done, namely; GRB 970508, GRB 980703, GRB 990510, GRB 991208 and GRB 991216. We compare our results to previous, multi-wavelength, modelling efforts. In Table~\ref{tab:grb_sample}, we present the overall properties of the GRB sample. For this work, we were able to generate \texttt{scalefit} tables for constant density CBM only. Therefore, constant density CBM is assumed when fitting the afterglow data. We use the prior distributions presented in Table~\ref{tab:prior_sample} in our modelling efforts. The inferred parameter distributions for the long GRB sample can be seen in Figure~\ref{fig:res_violin} in the form of a violin plot.

\begin{table}
    \caption{Redshift ($z$), Galactic foreground extinction ($A_{V, \mathrm{MW}}$), rest-frame host galaxy extinction ($A_{V, \mathrm{host}}$), and the best fit extinction model for the host galaxy of the long GRB sample. MW denotes Milky Way type host extinction, and SMC denotes Small Magellanic Cloud type host extinction \citep{Pei1992}.}
    \label{tab:grb_sample}
    \begin{tabular} {l r r r r}
        \hline
        GRB name &  \multicolumn{1}{c}{$z$}  &  \multicolumn{1}{c}{$A_{V, \mathrm{MW}}$} & \multicolumn{1}{c}{$A_{V, \mathrm{host}}$} & \multicolumn{1}{c}{Host type}\\
        \hline
        970508 & 0.835 & $\sim 0$ & $\sim 0$ & N/A\\
        980703 & 0.966 & $0.1891$ & $0.90$ & MW\\
        990510 & 1.619 & $\sim 0$ & $0.22$ & SMC\\
        991208 & 0.706 & $0.0512$ & $0.80$ & MW\\
        991216 & 1.02 & $2.016$ & $\sim 0$ & N/A\\
        \hline
    \end{tabular}
    
\end{table}

\begin{table}
    \setlength{\tabcolsep}{5pt}
    \def\arraystretch{1.25}
    \caption{Assumed priors for modelling the long GRB sample.}
    \begin{tabular}{l|r}
    \hline
    \multicolumn{1}{l}{Parameter range} & \multicolumn{1}{r}{Prior distribution} \\
    \hline
    $0.01<\theta_0<1.6$ & log-uniform\\
    
    $10^{48}<E_{K, \mathrm{iso}}<10^{55}$ & log-uniform\\
    
    $10^{-4}<n_{0}<1000$ &log-uniform\\
    
    $0<\theta_{\mathrm{obs}} / \theta_0<2$ & uniform\\
    
    $1.0<p<3.0$& uniform\\
    
    $10^{-7}<\epsilon_B<0.50$& log-uniform\\
    
    $10^{-5}<\bar{\epsilon_e}<10$& log-uniform\\
    \hline
    \end{tabular}
    \label{tab:prior_sample}
\end{table}

\subsection{GRB 970508}

\label{sec:res_970508}
GRB 970508 exhibits an increase in optical flux at around $\sim 1$ day after the burst, and starts to decline as a power-law with time. \cite{Panaitescu2002} (PK02 from now on) explain the rise in the optical flux by assuming that the jet is viewed off-axis with $\theta_{\mathrm{obs}} \sim 4/3\theta_0$. They find that a wind-like CBM ($n \propto r^{-2}$) suits the observations best. On the other hand, \cite{Yost2003} (Y03 from now on) favor a constant density CBM in their analysis.

In our analysis, we exclude the data points before the peak of the rise in optical wavelengths, and fit the data points which obey the power-law behaviour. We take the observing angle as a free parameter, allowing viewing angles both larger and smaller than the opening angle.

Figure~\ref{fig:lc_gp_best_970508} shows the light curves for the inferred parameter distribution for the afterglow of GRB 970508 using GP regression. Table~\ref{tab:970508} shows the inferred parameter values. Assuming that the re-brightening in optical bands is due to late-time energy injection from the central engine, the energetics inferred from our modelling will overestimate the initial explosion energy. We find a wide opening angle, $0.74^{+0.52}_{-0.28}$ rad, which is consistent with Y03, who find an opening angle of $0.84^{+0.03}_{-0.03}$ rad. The inferred isotropic kinetic energy by the GP model is consistent with both PK02 and Y03, whereas the $\chi^2$ sampling infers a lower $E_{K, \mathrm{iso}}$ with small uncertainty. Both GP likelihood and $\chi^2$ sampling infer similar $p$ values of $\sim 2.4$, which is larger than what PK02 and Y03 found. 

\begin{table*}
\centering
    \caption{Fit results for GRB 970508. Results from both $\chi^2$ sampling and GP likelihood sampling ($\mathcal{GP}$) are presented. The uncertainties on the parameters represent the 95\% credible interval for columns $\chi^2$ and $\mathcal{GP}$. Columns PK02 and Y03 show results from \protect\cite{Panaitescu2002} and \protect\cite{Yost2003}, respectively. PK02 did not provide uncertainties on the parameters for this burst. The uncertainties for the Y03 results represent the 68.3\% credible interval. All the values taken from previous studies have been converted to the same units. The values have been corrected for our choice of $\xi_N = 0.1$ by multiplying ($E_{K, \mathrm{iso}}$, $n_0$) by 10 and dividing ($\epsilon_B$, $\bar{\epsilon}_e$) by 10 (see Section~\protect\ref{sec:reg}).}
    \label{tab:970508}
    \setlength{\tabcolsep}{5pt}
    \def\arraystretch{1.25}
    \begin{tabular} {l r r r r}
        \hline
        \multicolumn{5}{c}{\textbf{GRB 970508}} \\
        \hline
        \hline
        Parameter &  \multicolumn{1}{c}{$\chi^2$} & \multicolumn{1}{c}{$\mathcal{GP}$} & \multicolumn{1}{c}{PK02} & \multicolumn{1}{c}{Y03}\\
        \hline
        {\boldmath$\theta_0       $} & $0.5007^{+0.0063}_{-0.0055}$ &  $0.74^{+0.52}_{-0.28}$ & 0.32 & $0.84^{+0.030}_{-0.030}$\\
        
        {\boldmath$\log_{10}(E_{K, \mathrm{iso}, 53})$} & $0.370^{+0.027}_{-0.026}$ & $0.53^{+0.28}_{-0.28}$ & 0.597 & $0.56^{+0.011}_{-0.011}$\\
        
        {\boldmath$\log_{10}(n_0) $} & $0.407^{+0.028}_{-0.027}$ &  $0.80^{+0.90}_{-0.69}$ & 0.87 & $0.30^{+0.021}_{-0.045}$ \\
        
        {\boldmath$\theta_{\mathrm{obs}}/\theta_0$} & $0.6977^{+0.0055}_{-0.0074}$ & $0.750^{+0.093}_{-0.096}$ & 1.33 & 0\\
        
        {\boldmath$p              $} & $2.404^{+0.013}_{-0.013}$ & $2.39^{+0.10}_{-0.12}$ & 2.18 & $2.12^{+0.03}_{-0.008}$  \\
        
        {\boldmath$\log_{10}(\epsilon_B)$} & $-2.729^{+0.075}_{-0.077}$ &  $-3.5^{+1.2}_{-1.4}$ & -2.34 & $-1.60^{+0.010}_{-0.036}$ \\
        
        {\boldmath$\log_{10}(\bar{\epsilon}_e)$} & $-2.078^{+0.011}_{-0.011}$ & $-1.97^{+0.30}_{-0.25}$ & -2.77 & $-2.43^{+0.02}_{-0.015}$\\
        \hline
    \end{tabular}
\end{table*}

\subsection{GRB 980703}

\cite{Panaitescu2001} (PK01 from now on) favor a constant density CBM for GRB 980703. Following \cite{Vreeswijk1999}, PK01 take the host extinction to be $A_V=1.45\pm0.13$, and find that $p=3.08$. Y03 also favor a constant density CBM, and find a value of $A_V=1.15$ for the host extinction and infer $p=2.54^{+0.04}_{-0.1}$. In this work we take the host extinction to be $A_V=0.9$ \citep{Bloom1998} and find that $p=2.05^{+0.10}_{-0.095}$.  

The host galaxy of GRB 980703 has a significant contribution to the observed radio and optical emission. We assume that the host galaxy contribution is constant over time and leave the host galaxy flux in radio and optical wavelengths as free parameters.

Figure~\ref{fig:lc_gp_best_980703} shows the light curves for the inferred parameter distribution for the afterglow of GRB 980703 using GP regression. Table~\ref{tab:980703} shows the inferred parameter values. The GP model infers an opening angle which is consistent with Y03, whereas $\chi^2$ likelihood sampling infers a smaller opening angle. Inferred $n_0$ and $E_{K, \mathrm{iso}}$ values are significantly smaller than Y03.
\begin{table}
% \centering
    \caption{Fit results for GRB 980703. Results from both $\chi^2$ sampling and GP likelihood sampling ($\mathcal{GP}$) are presented. See Table~\ref{tab:970508} for detailed explanation.}
    \label{tab:980703}
    \setlength{\tabcolsep}{5pt}
    \def\arraystretch{1.25}
    \begin{tabular} {l r r r}
        \hline
        \multicolumn{4}{c}{\textbf{GRB 980703}} \\
        \hline
        \hline
        Parameter &  \multicolumn{1}{c}{$\chi^2$} & \multicolumn{1}{c}{$\mathcal{GP}$} & \multicolumn{1}{c}{Y03}\\
        \hline
        {\boldmath$\theta_0       $} & $0.14990^{+0.00093}_{-0.0011}$ &  $0.199^{+0.043}_{-0.042}$ & $0.234^{+0.02}_{-0.007}$\\
        
        {\boldmath$\log_{10}(E_{K, \mathrm{iso}, 53})$} & $0.139^{+0.016}_{-0.015}$ & $-0.02^{+0.19}_{-0.17}$ & $1.07^{+0.028}_{-0.080}$\\
        
        {\boldmath$\log_{10}(n_0) $} & $0.156^{+0.026}_{-0.025}$ &   $0.58^{+0.33}_{-0.34}$  & $2.44^{+0.057}_{-0.049}$\\
        
        {\boldmath$\theta_{\mathrm{obs}}/\theta_0$} & $0.313^{+0.023}_{-0.018}$ & $0.31^{+0.28}_{-0.29}$ & 0\\
        
        {\boldmath$p              $} & $2.202^{+0.026}_{-0.026}$ & $2.049^{+0.092}_{-0.094}$ & $2.54^{+0.04}_{-0.1}$  \\
        
        {\boldmath$\log_{10}(\epsilon_B)$} & $-0.588^{+0.048}_{-0.049}$ &  $-0.87^{+0.28}_{-0.31}$ & $-3.74^{+0.087}_{-0.079}$\\
        
        {\boldmath$\log_{10}(\bar{\epsilon}_e)$} &$-2.520^{+0.015}_{-0.015}$ & $-2.36^{+0.13}_{-0.13}$ & $-2.02^{+0.065}_{-0.110}$ \\
        \hline
    \end{tabular}
\end{table}

\subsection{GRB 990510}

PK01 favor a constant density CBM for the case of GRB 990510. The optical afterglow of GRB 990510 exhibits a break in its temporal evolution at around 1.5 days. This break is interpreted as a jet-break. PK01 find $p=2.09 \pm 0.03$ using closure relations, which is also consistent with the inferred value from the GP model.

Figure~\ref{fig:lc_gp_best_990510} shows the light curves for the inferred parameter distribution for the afterglow of GRB 990510 using GP regression. Table~\ref{tab:990510} shows the inferred parameter values. The inferred opening angle is consistent with \cite{boxfit}, where they performed a detailed fit using \texttt{boxfit} (vE12) model and found an opening angle $0.075^{+0.002}_{-0.004}$ rad assuming an on-axis observer. On the other hand PK02 find a smaller opening angle $0.054^{+0.001}_{-0.006}$ rad. The GP model predicts a larger $\epsilon_B$ value when compared to $\chi^2$ sampling and previous studies.

\begin{table*}
\centering
    \caption{Fit results for GRB 990510. Results from both $\chi^2$ sampling and GP likelihood sampling ($\mathcal{GP}$) are presented. The uncertainties on the parameters represent the 95\% credible interval for columns $\chi^2$ and $\mathcal{GP}$. Columns PK02 and vE12 show results from \protect\cite{Panaitescu2002} and \protect\cite{boxfit}. The uncertainties for the PK02 and vE12 results represent the 90\% and 68\% credible intervals, respectively. All the values taken from previous studies have been converted to the same units. The values have been corrected for our choice of $\xi_N = 0.1$ by multiplying ($E_{K, \mathrm{iso}}$, $n_0$) by 10 and dividing ($\epsilon_B$, $\bar{\epsilon}_e$) by 10 (see Section~\protect\ref{sec:reg}). Since PK02 find $p < 2$, the conversion from $\epsilon_e$ to the equivalent $\bar{\epsilon}_e$ results in a negative value. Therefore we denote $\log_{10}(\bar{\epsilon}_e)$ as N/A.}
    \label{tab:990510}
    \setlength{\tabcolsep}{5pt}
    \def\arraystretch{1.25}
    \begin{tabular} {l r r r r}
        \hline
        \multicolumn{5}{c}{\textbf{GRB 990510}} \\
        \hline
        \hline
        Parameter &  \multicolumn{1}{c}{$\chi^2$} & \multicolumn{1}{c}{$\mathcal{GP}$} & \multicolumn{1}{c}{PK02} & \multicolumn{1}{c}{vE12}\\
        \hline
        {\boldmath$\theta_0       $} & $0.06423^{+0.00078}_{-0.00077}$ &  $0.0671^{+0.0062}_{-0.0042}$ & $0.054^{+0.0017}_{-0.0087}$ & $0.075^{+0.002}_{-0.004}$\\
        
        {\boldmath$\log_{10}(E_{K, \mathrm{iso}, 53})$} & $1.125^{+0.026}_{-0.024}$ &  $0.870^{+0.18}_{-0.089}$ & $0.98^{+0.80}_{-0.21}$ & $1.25^{+0.06}_{-0.02}$\\
        
        {\boldmath$\log_{10}(n_0) $} & $0.293^{+0.049}_{-0.042}$ &    $-0.01^{+0.15}_{-0.13}$ & $0.46^{+0.139}_{-0.316}$& $-0.52^{+0.054}_{-0.221}$\\
        
        {\boldmath$\theta_{\mathrm{obs}}/\theta_0$} & $0.705^{+0.014}_{-0.0093}$ & $0.297^{+0.057}_{-0.074}$ & 0 & 0\\
        
        {\boldmath$p              $} & $2.230^{+0.028}_{-0.027}$ & $1.91^{+0.12}_{-0.089}$ & $1.83^{+0.18}_{-0.01}$  & $2.28^{+0.06}_{-0.01}$\\
        
        {\boldmath$\log_{10}(\epsilon_B)$} & $-3.620^{+0.086}_{-0.089}$ &  $-1.72^{+0.42}_{-1.0}$ & $-3.28^{+0.95}_{-1.01}$& $-3.33^{+0.07}_{-0.08}$\\
        
        {\boldmath$\log_{10}(\bar{\epsilon}_e)$} &$-1.803^{+0.026}_{-0.026}$ & $-2.49^{+0.11}_{-0.12}$ & N/A& $-2.08^{+0.13}_{-0.02}$\\
        \hline
    \end{tabular}
\end{table*}

\subsection{GRB 991208}

Figure~\ref{fig:lc_gp_best_991208} shows the light curves for the inferred parameter distribution for the afterglow of GRB 991208 using GP regression. Table~\ref{tab:991208} shows the inferred parameter values. The inferred $p$ value by the GP model agrees with the results presented in PK02, whereas $\chi^2$ sampling results in a smaller $p$ value.

In our analysis we find a smaller opening angle than PK02 with an extremely off-axis observer angle. The GP regression and $\chi^2$ sampling give significantly different results for microphysical parameters and observer angle.
\begin{table}
% \centering
    \caption{Fit results for GRB 991208. Results from both $\chi^2$ sampling and GP likelihood sampling ($\mathcal{GP}$) are presented together with literature values. See Table~\ref{tab:990510} for detailed explanation.}
    \label{tab:991208}
    \setlength{\tabcolsep}{5pt}
    \def\arraystretch{1.25}
    \begin{tabular} {l r r r}
        \hline
        \multicolumn{4}{c}{\textbf{GRB 991208}} \\
        \hline
        \hline
        Parameter &  \multicolumn{1}{c}{$\chi^2$} & \multicolumn{1}{c}{$\mathcal{GP}$} & \multicolumn{1}{c}{PK02}\\
        \hline
        {\boldmath$\theta_0       $} & $0.02261^{+0.00059}_{-0.00059}$ &  $0.0350^{+0.0072}_{-0.0088}$ & $0.22^{+0.026}_{-0.038}$\\
        
        {\boldmath$\log_{10}(E_{K, \mathrm{iso}, 53})$} & $1.951^{+0.024}_{-0.024}$ &  $1.58^{+0.30}_{-0.23}$ & $-0.015^{+0.49}_{-0.27}$\\
        
        {\boldmath$\log_{10}(n_0) $} & $-0.1410^{+0.0073}_{-0.0077}$ &    $0.14^{+0.36}_{-0.32}$ & $2.25^{+0.34}_{-0.17}$\\
        
        {\boldmath$\theta_{\mathrm{obs}}/\theta_0$} & $0.0079^{+0.017}_{-0.0083}$ & $1.58^{+0.38}_{-0.34}$ & 0 \\
        
        {\boldmath$p              $} &$1.1964^{+0.0077}_{-0.0074}$& $1.55^{+0.13}_{-0.12}$ & $1.53^{+0.03}_{-0.03}$ \\
        
        {\boldmath$\log_{10}(\epsilon_B)$} & $-0.30186^{+0.00086}_{-0.0018}$ &  $-0.86^{+0.35}_{-0.38}$ & $-2.45^{+0.43}_{-0.39}$ \\
        
        {\boldmath$\log_{10}(\bar{\epsilon}_e)$} &$-3.241^{+0.017}_{-0.016}$ & $-2.57^{+0.23}_{-0.21}$ & N/A \\
        \hline
    \end{tabular}
\end{table}

\subsection{GRB 991216}

Figure~\ref{fig:lc_gp_best_991216} shows the light curves for the inferred parameter distribution for the afterglow of GRB 991216 using GP regression. Table~\ref{tab:991216} shows the inferred parameter values. PK02 find a hard electron distribution with $p = 1.36 \pm 0.03$, which is consistent with the results we get from GP modelling.

GP regression and $\chi^2$ sampling result in very different parameter values for GRB 991216. This is mainly because the optical data contribute to the $\chi^2$ value the most, whereas the radio data has a small contribution to the $\chi^2$. The best fit obtained from $\chi^2$ sampling, despite the fact that it has a smaller $\chi^2$ value than the best fit of the GP model, completely misses the radio data points and therefore is an inadequate representation of the observed emission.

\begin{table}
% \centering
    \caption{Fit results for GRB 991216. Results from both $\chi^2$ sampling and GP likelihood sampling ($\mathcal{GP}$) are presented together with literature values. See Table~\ref{tab:990510} for detailed explanation.}
    \label{tab:991216}
    \setlength{\tabcolsep}{5pt}
    \def\arraystretch{1.25}
    \begin{tabular} {l r r r}
        \hline
        \multicolumn{4}{c}{\textbf{GRB 991216}} \\
        \hline
        \hline
        Parameter &  \multicolumn{1}{c}{$\chi^2$} & \multicolumn{1}{c}{$\mathcal{GP}$} & \multicolumn{1}{c}{PK02} \\
        \hline
        {\boldmath$\theta_0       $} &$0.1268^{+0.0031}_{-0.0095}$ & $0.033^{+0.028}_{-0.012}$ & $ 0.047^{+0.006}_{-0.017}$ \\
        
        {\boldmath$\log_{10}(E_{K, \mathrm{iso}, 53})$} & $1.9961^{+0.0041}_{-0.0082}$ & $1.34^{+0.36}_{-0.51}$ & $ 0.99^{+0.68}_{-0.31}$  \\
        
        {\boldmath$\log_{10}(n_0) $} & $-0.6833^{+0.0094}_{-0.010}$ & $-0.60^{+1.0}_{-0.64}$ & $ 1.67^{+0.38}_{-0.20}$  \\
        
        {\boldmath$\theta_{\mathrm{obs}}/\theta_0$} & $0.7979^{+0.0071}_{-0.0095}$ & $0.43^{+0.54}_{-0.33}$ & 0  \\
        
        {\boldmath$p              $} & $2.536^{+0.016}_{-0.017}$ & $1.49^{+0.18}_{-0.16}$  & $ 1.36^{+0.03}_{-0.03}$ \\
        
        {\boldmath$\log_{10}(\epsilon_B)$} & $-5.041^{+0.044}_{-0.036}$ & $-0.76^{+0.46}_{-0.70}$  & $ -2.74^{+0.46}_{-0.21}$ \\
        
        {\boldmath$\log_{10}(\bar{\epsilon}_e)$} & $-1.601^{+0.025}_{-0.028}$ & $-2.77^{+0.42}_{-0.36}$  & N/A \\
        \hline
    \end{tabular}
\end{table}

\section{Discussion}
\label{sec:discuss}
GRBs are thought to be collimated outflows, therefore the isotropic equivalent energies of these events are an overestimation of the true energetics. The true, beaming corrected, energies of these events significantly depend on the geometry of the outflow (i.e. the opening angle),

\begin{equation}
    E_K = E_{K, \mathrm{iso}}(1 - \cos\theta_0).
\end{equation}

\noindent Previous studies have shown that there is observational evidence that there exists a standard energy reservoir for GRBs. \cite{Frail2001} have measured the opening angle of a sample of GRBs based on achromatic breaks in the afterglow light curve. They have shown that the beaming corrected energy release in $\gamma$-rays is narrowly clustered around $5\times10^{50}$ erg. Moreover, \cite{Panaitescu2002} have shown, using multi-wavelength afterglow modelling, that the beaming corrected kinetic energy of GRBs are narrowly distributed and vary between $10^{50}$ to $5 \times  10^{50}$ erg. Similarly, \cite{Berger2003} have analysed the X-ray afterglow data for a large sample of GRBs with known jet breaks, and have found evidence that the beaming corrected kinetic energy of these events are approximately constant. Note that, these studies are based on pre-\textit{Swift} afterglow observations and therefore might be biased towards more energetic GRBs. After the launch of \textit{Swift}, thanks to improved localization, the number of redshift measurements have increased and other classes of GRBs have been discovered, such as low-luminosity GRBs (ll-GRBs) which exhibit significantly less luminous prompt emission. ll-GRBs might constitute outliers in the overall GRB population and might not conform with the idea of a constant energy reservoir \citep{Liang2007}.

\begin{figure}
 \includegraphics[width=\columnwidth]{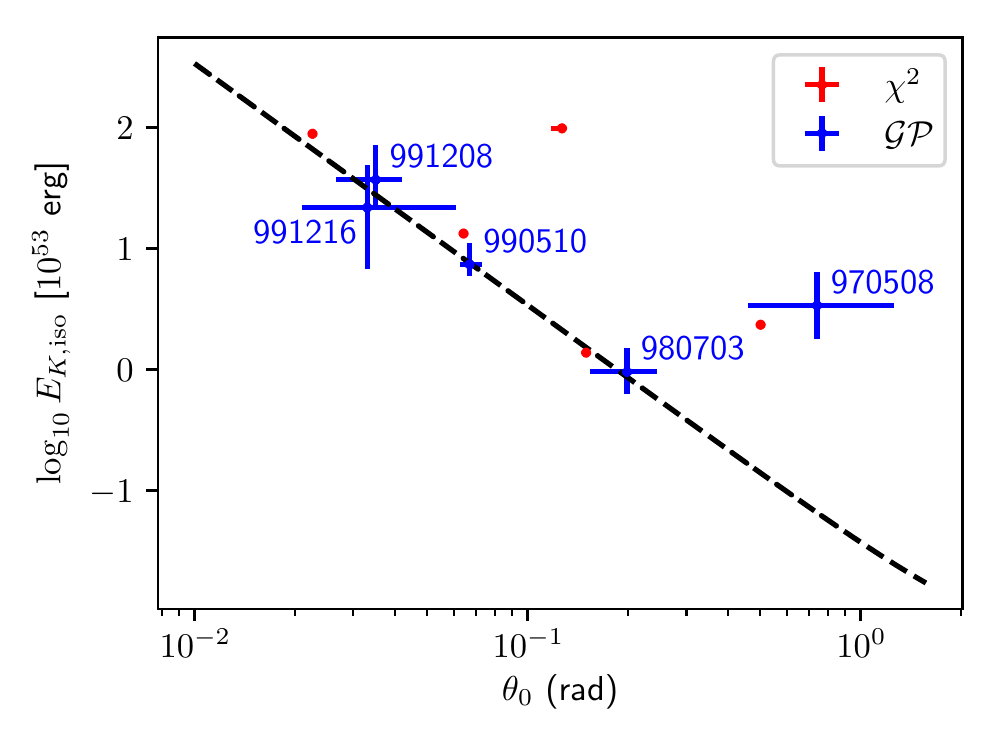}
 \caption{Isotropic equivalent kinetic energy ($E_{K\mathrm{iso}}$) dependence on the opening angle ($\theta_0$) inferred from our modelling. The red measurements ($\chi^2$) are obtained by $\chi^2$ sampling and the blue measurements ($\mathcal{GP}$) are obtained by sampling the GP log likelihood function. The dashed black line represents the $E_{K,\mathrm{iso}}(1 - \cos{\theta_0}) = 1.7\times10^{51}$ relation. The error bars represent the 95\% credible interval.}
 \label{fig:E_Kiso_theta_0}
\end{figure}

Our analysis also shows a strong correlation between $\theta_0$ and $E_{K, \mathrm{iso}}$. In Figure~\ref{fig:E_Kiso_theta_0}, we show the measured opening angles and isotropic energies of our GRB sample. It can be seen that, when the GP model is used for inferring parameters, the measured values suggest that the beaming corrected kinetic energies are approximately the same for long GRBs. GRB 970508 is a clear outlier, which is consistent with the findings of \cite{Panaitescu2002}. As discussed in Section~\ref{sec:res_970508}, GRB 970508 exhibits a re-brightening in optical wavelengths, which could be due to late-time energy injection. This energy injection could account for the overestimation of the isotropic-equivalent kinetic energy of this source, which might imply that the standard energy reservoir applies more strongly to the initial ejecta formation of a GRB than to any later activity of the central engine. Note that the correlation is not apparent when $\chi^2$ sampling is used for parameter estimation.

\begin{figure}
 \includegraphics[width=\columnwidth]{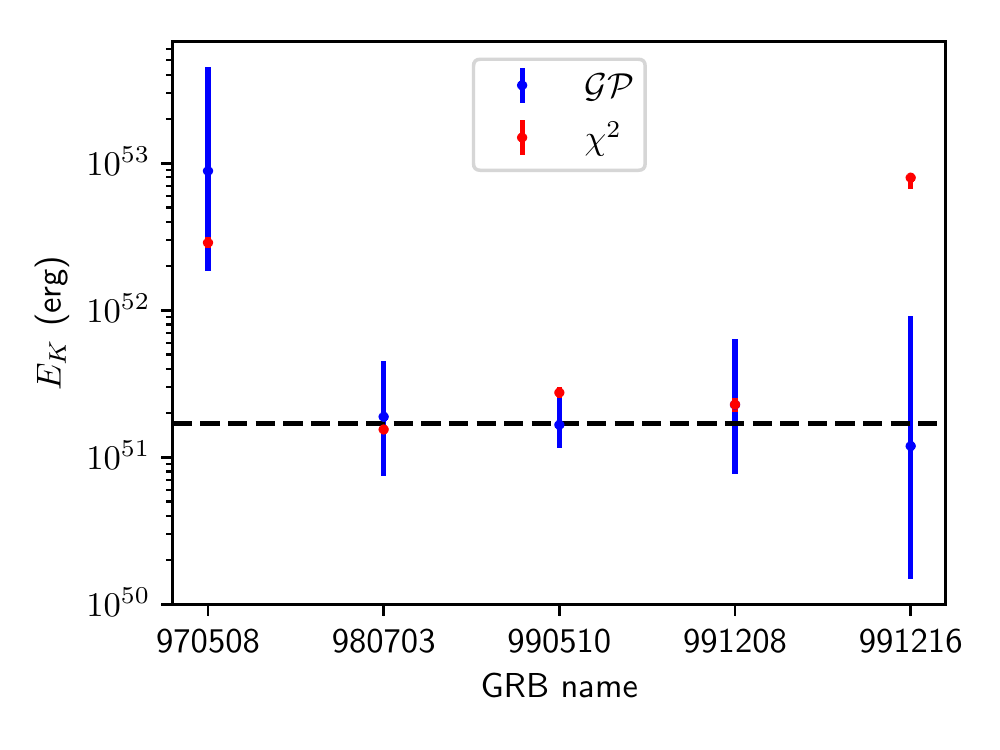}
 \caption{Beaming corrected kinetic energies of the long GRB sample. The red measurements ($\chi^2$) are obtained by $\chi^2$ sampling and the blue measurements ($\mathcal{GP}$) are obtained by sampling the GP log likelihood function. The dashed line is the log-average of GRBs 980703, 990510, 991208, 991216, which is equal to $1.7\times10^{51}$ erg. The error bars represent the 95\% credible interval.}
 \label{fig:E_K_corrected}
\end{figure}

In Figure~\ref{fig:E_K_corrected}, we show the inferred beaming corrected kinetic energies with 95\% credible intervals for the sample GRBs. The inferred $E_K$ in our analysis is $\sim 1.7 \times 10^{51}$ erg, which is about an order of magnitude larger than what previous studies have found. This discrepancy with previous studies is expected as we fix $\xi_N$ to be 0.1 instead of the canonical value of 1.0. We also recognise that it is too early to judge whether these few very well studied, well-sampled GRB afterglows are representative of the whole population.

\section{Conclusion}
\label{sec:conclude}
In this work, we have introduced a novel method for modelling GRB afterglows, where Gaussian processes are used to take into account any systematics between the model and observations in a non-parametric fashion. Using synthetic data sets, we have shown that the GP approach results in more accurate posterior distributions with respect to sampling the $\chi^2$ likelihood.

We model a sample of 5 well-known long GRBs with multi-wavelength coverage (GRBs 970508, 980703, 990510, 991208, 991216), using the \texttt{scalefit} code together with the GP framework. We compare the inferred  parameters for each GRB with the literature values and comment upon the parameter distributions of the overall sample. We find a correlation between the isotropic-kinetic energy and opening angle, with GRB 970508 being the only outlier. This correlation, which is consistent with previous studies, suggests that there is a common energy reservoir which drives the dynamics of GRBs.

\section*{Acknowledgements}
We would like to thank Vatsal Panwar for valuable discussions on Gaussian processes. We would like to thank Dr. Elena Sellentin for her useful tips regarding coverage measurements. We would also like to thank the anonymous referee for their detailed comments. H. J. van Eerten acknowledges partial support by the European Union Horizon 2020 Programme under the AHEAD2020 project (grant agreement number 871158). This work was sponsored by NWO Exact and Natural Sciences for the use of supercomputer facilities.

\section*{Data Availability}

The data underlying this article can be shared on reasonable request to the corresponding author.
% The Acknowledgements section is not numbered. Here you can thank helpful
% colleagues, acknowledge funding agencies, telescopes and facilities used etc.

%%%%%%%%%%%%%%%%%%%%%%%%%%%%%%%%%%%%%%%%%%%%%%%%%%



\bibliographystyle{mnras}
\bibliography{references} % if your bibtex file is called example.bib

% \begin{thebibliography}{99}
% \bibitem[\protect\citeauthoryear{Author}{2012}]{Author2012}
% Author A.~N., 2013, Journal of Improbable Astronomy, 1, 1
% \bibitem[\protect\citeauthoryear{Others}{2013}]{Others2013}
% Others S., 2012, Journal of Interesting Stuff, 17, 198
% \end{thebibliography}

%%%%%%%%%%%%%%%%%%%%%%%%%%%%%%%%%%%%%%%%%%%%%%%%%%

%%%%%%%%%%%%%%%%% APPENDICES %%%%%%%%%%%%%%%%%%%%%
% \clearpage
\appendix

\section{Fit results }

\begin{figure}
 \includegraphics[width=\columnwidth]{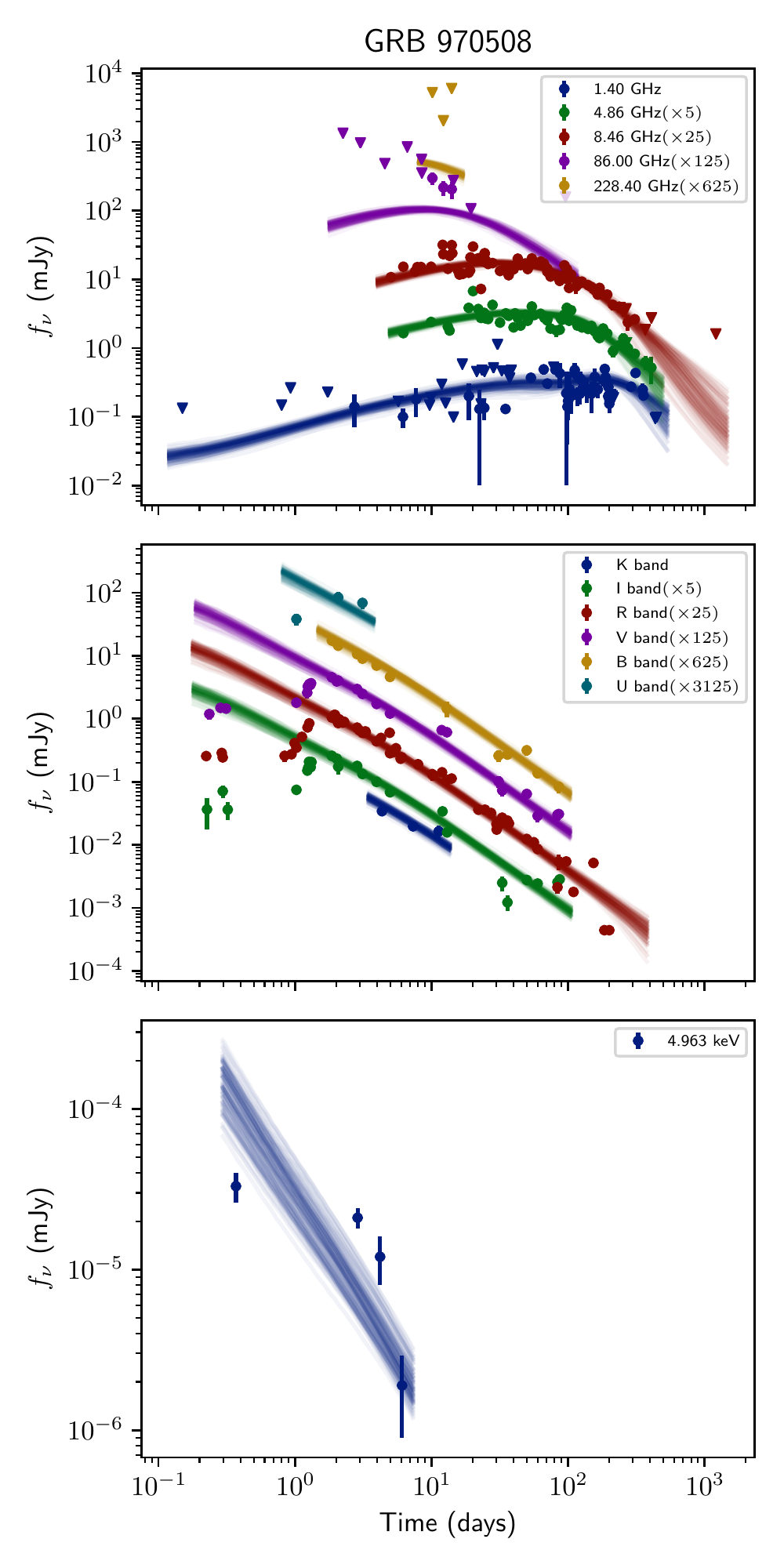}
 \caption{Fit result for GRB 970508 by sampling the GP likelihood. Observed flux density values are presented in radio, optical and X-ray bands (upper, middle and lower panel respectively) together with the posterior predictive light curves. Triangles represent 3-$\sigma$ upper limits.} A sample of 100 parameter sets are randomly drawn from the inferred joint probability distribution of the parameters, and \texttt{scalefit} light curves are drawn for each parameter set.
 \label{fig:lc_gp_best_970508}
\end{figure}

\begin{figure}
 \includegraphics[width=\columnwidth]{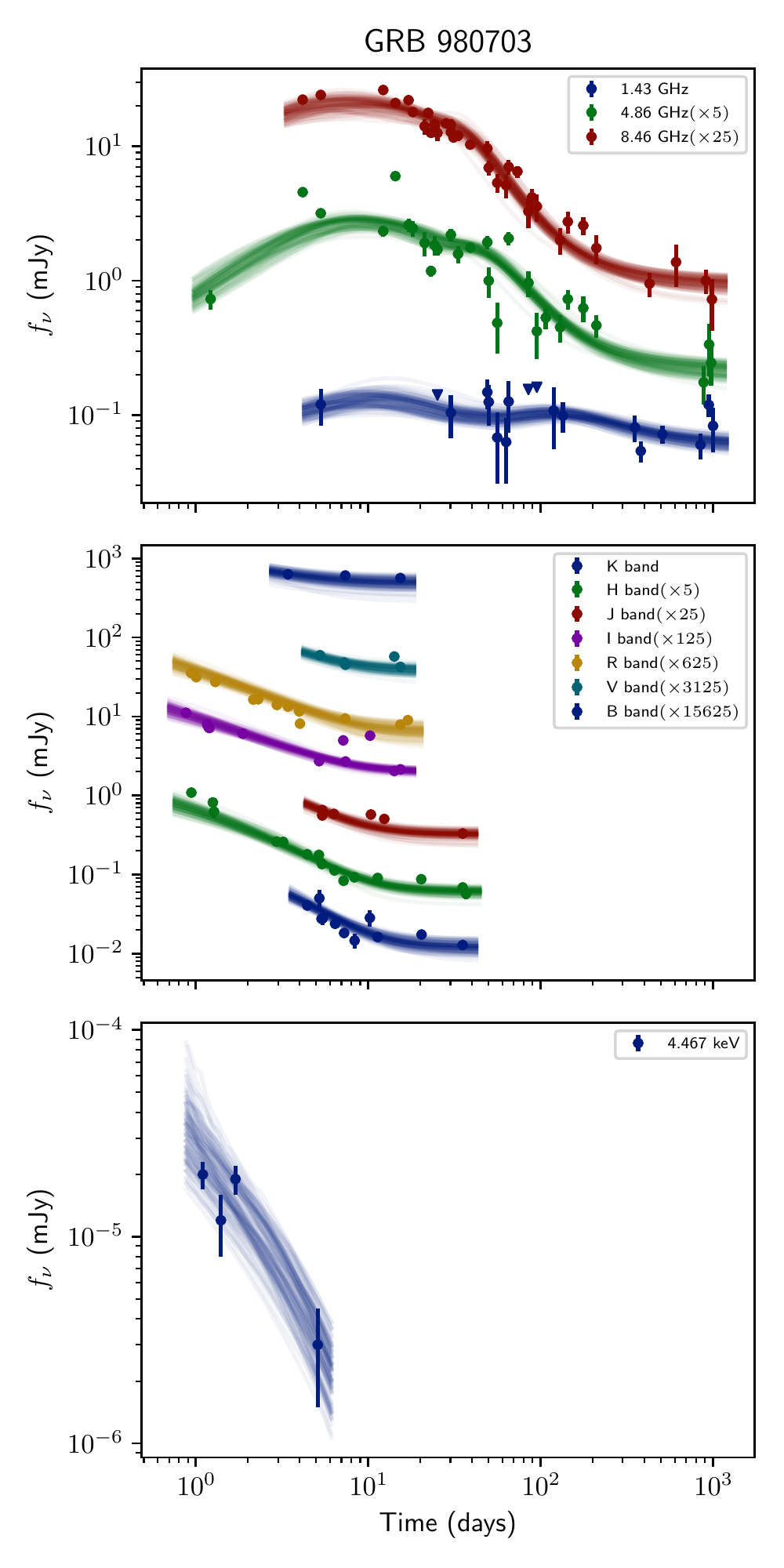}
 \caption{Fit result for GRB 980703 by sampling the GP likelihood. Observed flux density values are presented in radio, optical and X-ray bands (upper, middle and lower panel respectively) together with the posterior predictive light curves. Triangles represent 3-$\sigma$ upper limits.} A sample of 100 parameter sets are randomly drawn from the inferred joint probability distribution of the parameters, and \texttt{scalefit} light curves are drawn for each parameter set. The host galaxy contribution in radio and optical is not subtracted.
 \label{fig:lc_gp_best_980703}
\end{figure}

\begin{figure}
 \includegraphics[width=\columnwidth]{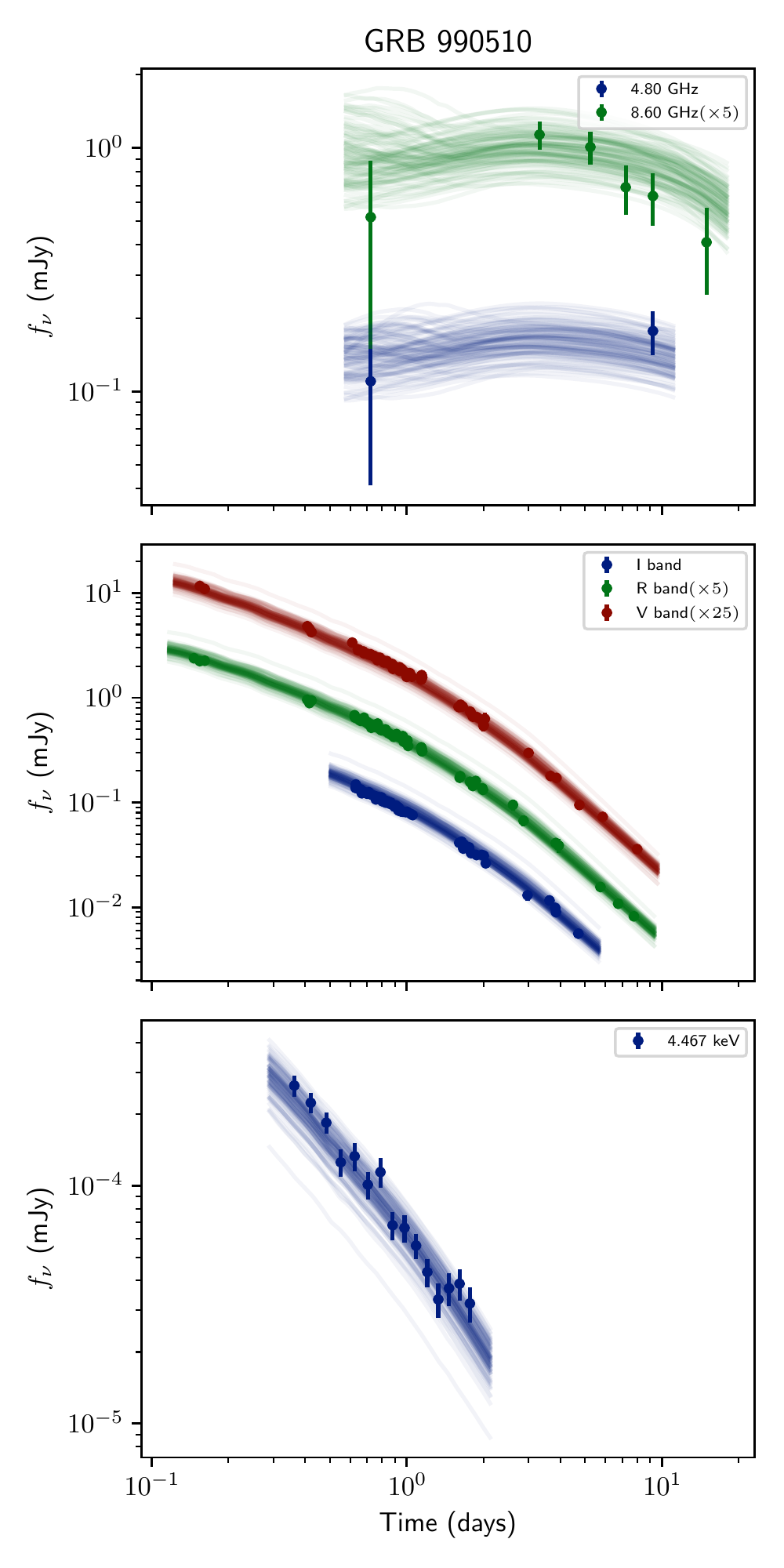}
 \caption{Fit result for GRB 990510 by sampling the GP likelihood. Observed flux density values are presented in radio, optical and X-ray bands (upper, middle and lower panel respectively) together with the posterior predictive light curves.} A sample of 100 parameter sets are randomly drawn from the inferred joint probability distribution of the parameters, and \texttt{scalefit} light curves are drawn for each parameter set.
 \label{fig:lc_gp_best_990510}
\end{figure}

\begin{figure}
 \includegraphics[width=\columnwidth]{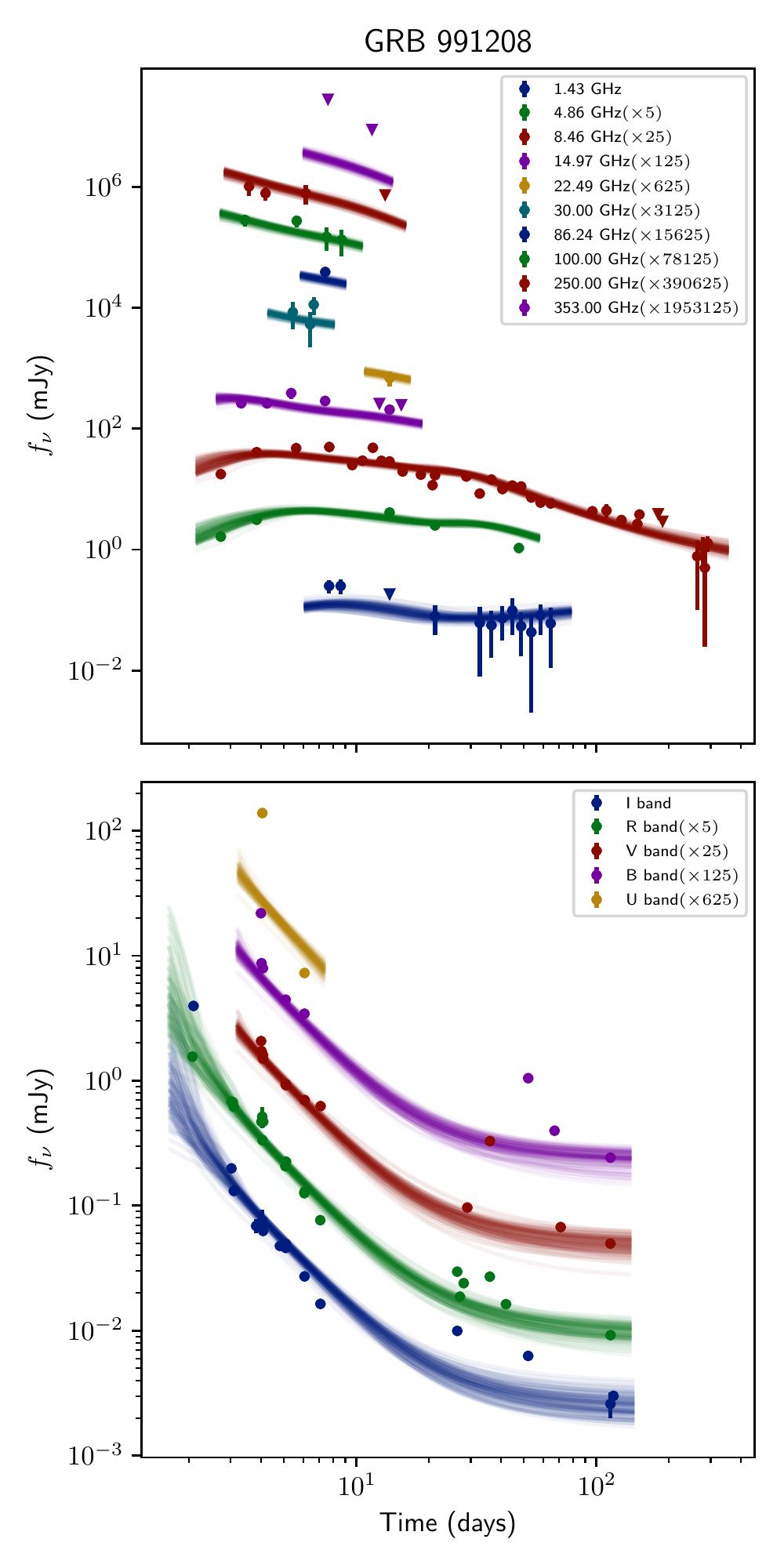}
 \caption{Fit result for GRB 991208 by sampling the GP likelihood. Observed flux density values are presented in radio and optical bands (upper and lower panel respectively) together with the posterior predictive light curves. Triangles represent 3-$\sigma$ upper limits.} A sample of 100 parameter sets are randomly drawn from the inferred joint probability distribution of the parameters, and \texttt{scalefit} light curves are drawn for each parameter set. The host galaxy contribution in optical is not subtracted.
 \label{fig:lc_gp_best_991208}
\end{figure}

\begin{figure}
 \includegraphics[width=\columnwidth]{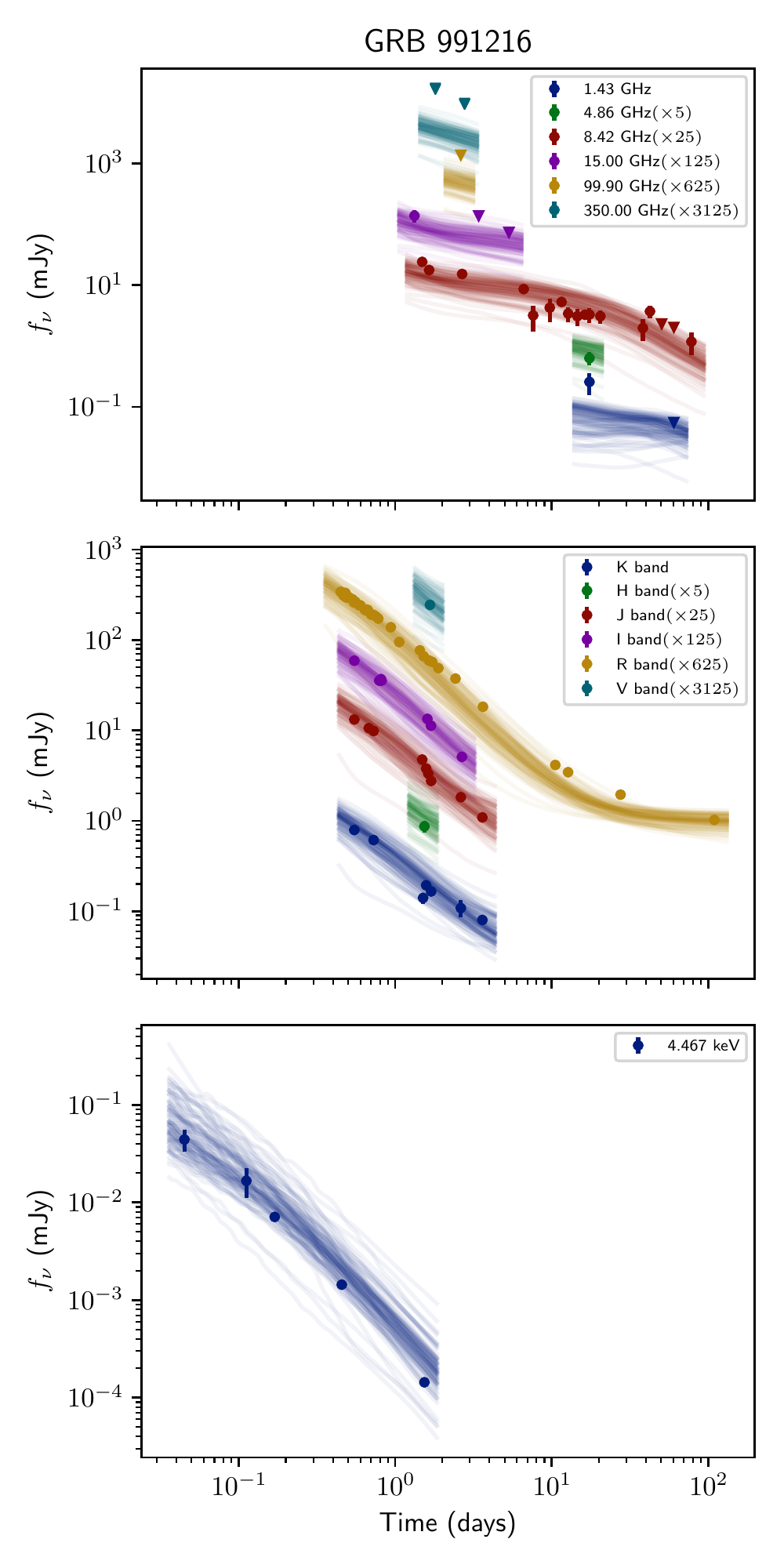}
 \caption{Fit result for GRB 991216 by sampling the GP likelihood. Observed flux density values are presented in radio, optical and X-ray bands (upper, middle and lower panel respectively) together with the posterior predictive light curves. Triangles represent 3-$\sigma$ upper limits.} A sample of 100 parameter sets are randomly drawn from the inferred joint probability distribution of the parameters, and \texttt{scalefit} light curves are drawn for each parameter set. The host galaxy contribution in optical is not subtracted.
 \label{fig:lc_gp_best_991216}
\end{figure}

% If you want to present additional material which would interrupt the flow of the main paper,
% it can be placed in an Appendix which appears after the list of references.

%%%%%%%%%%%%%%%%%%%%%%%%%%%%%%%%%%%%%%%%%%%%%%%%%%

% Don't change these lines
\bsp	% typesetting comment
\label{lastpage}
\end{document}